\documentclass[sigconf,10pt,nonacm]{acmart}

\usepackage[english]{babel}
\usepackage{graphicx} %
\usepackage{subcaption}
\usepackage{gensymb}

\newcommand*{\eg}{\textit{e.g.,}~}
\newcommand*{\ie}{\textit{i.e.,}~}

\newcommand{\squishlist}{
 \begin{list}{${\bullet}$}
  { \setlength{\itemsep}{0pt}
     \setlength{\parsep}{1pt}
     \setlength{\topsep}{1pt}
     \setlength{\partopsep}{0pt}
     \setlength{\leftmargin}{1em}
     \setlength{\labelwidth}{0.5em}
     \setlength{\labelsep}{0.5em} } }

\newcommand{\squishend}{
  \end{list}  }

\title{Sustainability or Survivability? Eliminating the Need to Choose in LEO Satellite Constellations}

\author{Chris Misa}
\affiliation{%
  \institution{University of Oregon}
  \country{}
}

\author{Ramakrishnan Durairajan}
\affiliation{%
  \institution{University of Oregon and Link Oregon}
  \country{}
}

\begin{abstract}
LEO Satellite Networks (LSNs) are revolutionizing global connectivity, but their reliance on tens of thousands of satellites raises pressing concerns over sustainability and survivability. 
In this work, we argue that the inefficiencies in LSN designs stem from ignoring the strong spatiotemporal structure of Internet traffic demand (which impacts sustainability) and the physical realities of the near-Earth space environment (which affects survivability). 
We propose a novel design approach based on sun-synchronous (SS) orbits called SS-plane, which aligns satellite coverage with the Earth’s diurnal cycle. 
We demonstrate that SS-plane constellations can reduce the number of satellites required by up to an order of magnitude and cut radiation exposure by $\sim$23\% compared to traditional Walker-delta constellations. 
These findings suggest a paradigm shift in LSN research from large, disposable megaconstellations to more sustainable, targeted LEO constellations.
\end{abstract}

\begin{document}

\maketitle

\vspace{-0.2cm}
\section{Introduction}
\label{sec:intro}

Low Earth Orbit (LEO) Satellite Networks (LSNs) are quickly becoming a cornerstone of global Internet infrastructure, offering fast, reliable connectivity nearly anywhere on the planet~\cite{kassing2020hypatia,handley2018delay,reutersTmoStarlink,bhattacherjee2019network,lai2023starrynet}. 
These systems are especially attractive for their potential to close the digital divide by delivering low-latency direct-to-consumer and transit service to regions of the planet that are difficult to connect with terrestrial network technologies~\cite{zhao2024low,li2024satguard,lai2024spacertc,stevens2024can,starlinkTransit,telesatNoStarlink}. 

However, to meet both bandwidth demand and survivability goals, current LSNs require tens of thousands of satellites, raising serious concerns about their environmental impact and long-term sustainability~\cite{ferreira2024potential,walker2020impact,massiveRisks,gaoReport,fasArticle}.
In particular, the limited capacity of a single satellite's spot-beams implies LSNs must use overlapping coverage from many satellites to satisfy bandwidth demand of populated regions.
Meanwhile, the harsh radiation environment of near-Earth space, coupled with cost pressures to reduce shielding,
implies LSNs must operate extra in-orbit backup satellites to replace active satellites as they inevitably fail~\cite{lai2024your,fccSES02767}.
{\em Thus, there is a fundamental tension: reducing satellite count improves sustainability but jeopardizes survivability, while adding redundant satellites improves survivability but compromises sustainability.}

Recent studies~\cite{ferreira2024potential,liou2008instability,jang2022stability,d2022analysis} demonstrate dire environmental consequences of continuous launch and disposal in the upper atmosphere and inevitable instability resulting from overcrowding of limited physical LEO real estate if these pressures towards massive mega-constellations continue unabated.
In response, the networking community has begun to explore design strategies that optimize satellite usage rather than maximize it~\cite{lai2024your,chen2024unraveling}.
A key observation of these strategies is that the spatial demand for satellite networking is sparse and clustered across the Earth surface whereas the supply of present-day LSNs is more-or-less uniform.
LSNs could, in principle, be designed to provide non-uniform coverage to only the surface regions with higher demand, thereby reducing the number of satellites required.

Unfortunately, this purely {\em spatial} approach of designing more efficient LSNs is not feasible in practice.
The relatively high velocity of satellites at LEO altitudes causes them to make $O$(10) trips around the Earth each day, which (when combined with the natural rotation of the Earth w.r.t. satellite orbital planes) makes focusing LSN bandwidth supply in a spatially controlled and consistent manner nearly impossible.
In particular, we show that the recently proposed use of repeat ground-track (RGT) orbits~\cite{chen2024unraveling} performs strictly worse than present-day Walker-delta constellation designs in terms of the number of satellites required.

Instead of relying solely on spatial structure, our key insight is that LSN designs must also focus on {\em temporal} structure (i.e., the diurnal rhythms and seasonality of human Internet usage).
Network traffic, especially in access ISP or mobile provider scenarios commonly targeted by LSNs, %
exhibits strong daily cycles caused by waking and sleeping hours~\cite{shen2016follow,koumar2025cesnet,macia2018ugr}.
Because the seasonality of each particular demand location on the Earth surface is synchronized with the Earth's rotation on its axis, this implies {\em traffic demand is actually fixed in space relative to the sun.}

This key insight opens up a tantalizing opportunity to rethink the LSN constellation design. 
We argue that their massive size is not an inherent requirement of satellite networking but rather a symptom of architectures that ignore the spatiotemporal structure of global demand and orbital hazards. 
To address this, we propose designing LSNs from a {\em spatiotemporal perspective}, treating both human activity patterns and orbital dynamics as first-class design inputs.

To move beyond these abstract arguments, we introduce a new primitive for LSN design called {\em sun-synchronous plane} (henceforth {\em SS-plane}), based on sun-synchronous (SS) orbits~\cite{vallado2001fundamentals}. %
SS-plane constellations align satellite coverage with the solar-fixed rhythm of human demand, achieving more with fewer satellites. 
We show that SS-plane constellations can reduce satellite count by up to an order of magnitude compared to Walker-delta designs. 
They also expose satellites to significantly less radiation, lowering average exposure by up to 23\%. 
Finally, we explore the implications of this approach for LSN research such as topology design, routing, and traffic control.

What makes this work both timely and provocative is that we challenge the prevailing ``more is better" philosophy behind today's LEO megaconstellations. 
By showing that it is possible to dramatically reduce satellite count and radiation exposure through sun-relative, demand-aware designs, we question the %
necessity of this design philosophy and its concomitant survivability and sustainability challenges. 
In doing so, 
we introduce a fundamentally different vision to rethink coverage not in {\em geo}centric but in {\em helio}centric terms, aligning network infrastructure with human activity and refocusing LSN research on synchronizing supply and demand to the spinning Earth, not just to points fixed on its surface. %

\vspace{-0.2cm}
\section{Background \& Motivation}

\subsection{Why are present-day LSNs so big?}

Present-day %
LSNs are immense in scale, often comprising thousands or even tens of thousands of satellites. 
For example, SpaceX recently announced plans to expand Starlink to nearly 30k satellites~\cite{starlink24art,starlink24art2}.
This scale is primarily driven by two operational necessities: the need to meet growing bandwidth demands and the requirement for survivability in the face of frequent hardware failures. 

In terms of bandwidth, a single satellite's spot-beam capacity is limited. That is, to support dense population centers with high data usage, overlapping coverage from multiple satellites is necessary~\cite{pucholModel}. 
This overlap inherently increases the number of satellites needed in orbit at any given time.

In terms of survivability, a chief hazard faced by LEO satellites is bombardment by various species of charged particles (\eg electrons, protons) trapped in the Earth's magnetic field which impact reliability and longevity of sensitive onboard networking equipment%
~\cite{misa2025reimagined,girgis2022radiation,guild2022best}. 
Since building highly shielded, fault-tolerant satellites would dramatically increase costs and mass, current systems opt instead to keep spare satellites in orbit (\eg 2-10 per orbital plane)~\cite{lai2024your,fccSES02767}. 
These spares can be quickly ``hot-swapped'' to replace failed units, but this redundancy further inflates the total satellite count. 
Thus, the prevailing approach in LSN design has treated satellites as cheap and disposable, reinforcing the trend toward massive constellations.

{\bf The LSN minimization challenge.}
This reality sets up the key design challenge for LSNs: {\em how to minimize the number of satellites without compromising the system's ability to meet bandwidth demand and maintain network survivability?} 

\subsection{Repeat ground-track orbits are no silver bullet}

Prior work~\cite{chen2024unraveling} attempted to address this challenge via more efficient constellation geometries with non-uniform spatial coverage, turning to repeat ground-track (RGT) orbits.
These orbits have their altitudes (and hence orbital periods) carefully tuned against the spin of the Earth so that they retrace the same path over Earth’s surface on a regular basis, making them attractive for applications requiring consistent coverage of specific regions
~\cite{vallado2001fundamentals}.

\begin{figure}[!htb]
    \centering
    \vspace{-0.1cm}
    \includegraphics[width=0.9\linewidth]{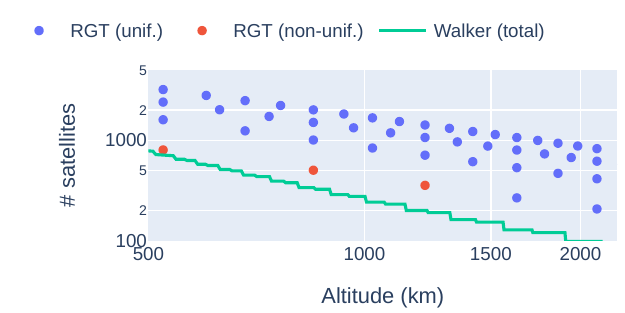}
    \caption{Minimum number of satellites required to cover a single repeat ground-track compared to number of satellites required for uniform global coverage in a Walker-delta constellation at 65$\degree$ inclination.}
    \vspace{-0.1cm}
    \label{fig:rgt_number_sats}
    \vspace{-0.1cm}
\end{figure}

However, to ensure continuous coverage over time, a single repeat ground-track from LEO (\eg using equally-space satellites with overlapping FOVs) requires more satellites than a minimal uniform coverage Walker-delta constellation at the same altitude as shown in Figure~\ref{fig:rgt_number_sats}.
For example, covering the RGT at 1215~km altitude requires $\geq$356 satellites compared to the $\geq$200 required for uniform global coverage in a Walker-delta constellation at the same altitude.

\begin{figure}[!htb]
    \centering
    \includegraphics[width=0.8\linewidth]{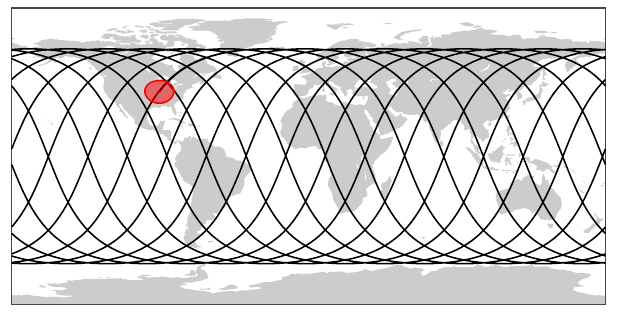}
    \caption{Example of a repeat ground-track with inclination 65$\degree$, altitude $\sim$560~km and the surface region coverage by a single satellite covering this ground-track (in red).}
    \label{fig:rgt_example}
\end{figure}

Moreover, at LEO altitudes the adjacent passes of the RGT are densely packet around the Earth's circumference (as illustrated in Figure~\ref{fig:rgt_example}), leading to near-uniform global coverage anyway.
Figure~\ref{fig:rgt_number_sats} confirms that {\em only three} of the possible RGTs at LEO {\em do not} automatically provide uniform global coverage.

{\bf Key takeaway.} The limitations of both traditional Walker-delta and RGT orbits expose a critical shortcoming in existing constellation strategies: a lack of alignment with the actual structure of user demand. Any effort to reduce the number of satellites meaningfully must account for where and when bandwidth is needed---not just how to cover the Earth uniformly. This insight motivates a shift toward demand-centric design principles that better reflect the spatiotemporal nature of global Internet usage.

More broadly, the inefficiency of current LSNs stems not just from their geometric design but from a more fundamental mismatch between satellite coverage (largely uniform) vs. human activity (and hence network traffic) is anything but uniform.

\section{Structural Considerations in the LSN Design Problem}

To move toward more sustainable and survivable constellations, we must examine the intrinsic structure of Internet demand as well as the physical constraints imposed by the near-Earth space environment respectively. We posit that these two factors---spatiotemporal demand and orbital radiation---form the foundation of any design to reduce satellite counts without sacrificing quality of service or reliability.

\subsection{The structure of bandwidth demand}
\label{ssec:structureOfBandwidth}

The primary driver behind designing efficient LSNs lies in understanding the structure of Internet bandwidth demand. Bandwidth usage is highly non-uniform across both space and time. 

{\bf Spatial structure.} Network bandwidth demand mirrors the highly non-uniform spatial distribution of human population across the Earth surface. Intermediate latitudes which are home to much of the world's population experience particularly high demand, while polar and oceanic regions see minimal usage. This skewed distribution suggests that uniform satellite coverage is an inefficient strategy, as it dedicates significant resources to areas with little or no demand.

\begin{figure}[!htb]
    \centering
    \includegraphics[width=0.9\linewidth]{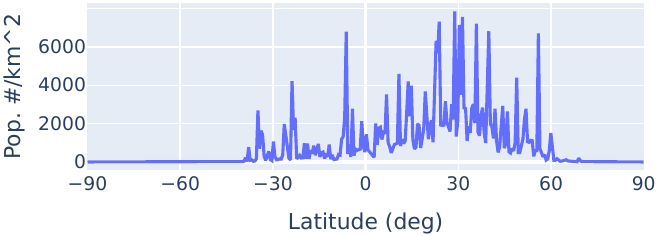}
    \caption{Maximum global population at each latitude aggregated in $0.5\degree$ bins (from~\cite{sedacGWP}).}
    \label{fig:gwpdLat}
\end{figure}

To illustrate we consider the SEDAC Gridded World Population Density~\cite{sedacGWP} data which estimates the population density of a 0.5$\degree$-wide grid over the Earth surface.
Figure~\ref{fig:gwpdLat} shows the maximum population density over all longitudes for each latitude indicating significant clustering of population at intermediate latitudes.

{\bf Temporal structure.} Temporally, Internet traffic driven by human activity exhibits pronounced daily cycles. Usage rises during waking hours and drops sharply at night. This diurnal pattern repeats reliably and is aligned with the Earth's rotation, meaning that peak usage in any given region is tied to local solar time.

\begin{figure}[!htb]
    \centering
    \includegraphics[width=0.9\linewidth]{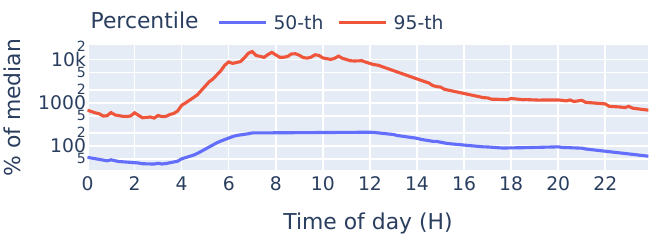}
    \caption{Bandwidth demand measured in bytes as a function of local time-of-day across 283 sites in CESNET~\cite{koumar2025cesnet} over a year.}
    \label{fig:cesnetExample}
\end{figure}

To illustrate, we consider the CESNET-TimeSeries24 dataset which includes fine-granularity throughput measurements from 283 sites across the Czech Republic~\cite{koumar2025cesnet}.
To summarize the daily seasonality in this dataset, we normalize throughput at each site by the site's median (to account for relative differences between sites) and group by time of day.
Figure~\ref{fig:cesnetExample} shows the resulting relationship between time of day and bandwidth demand (median and 95-percentile over all sites).

{\bf Spatiotemporal structure.} Intuitively, we can combine the data from Figure~\ref{fig:gwpdLat} and Figure~\ref{fig:cesnetExample} by scaling the bandwidth demand at each point in time by the population density at each point in space to approximate the spatiotemporal structure of bandwidth demand.

\begin{figure*}[!htb]
    \centering
    \hfill
    \begin{subfigure}{0.21\linewidth}
        \includegraphics[width=\linewidth]{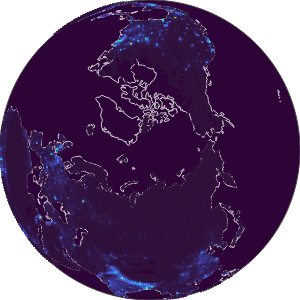}
        \caption{Hour 0}
    \end{subfigure}\hfill
    \begin{subfigure}{0.21\linewidth}
        \includegraphics[width=\linewidth]{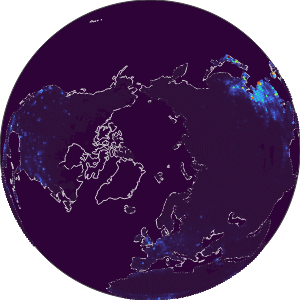}
        \caption{Hour 6}
    \end{subfigure}\hfill
    \begin{subfigure}{0.21\linewidth}
        \includegraphics[width=\linewidth]{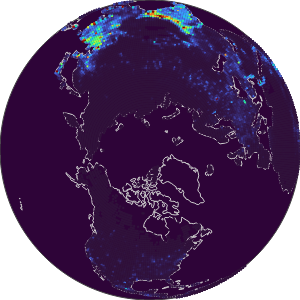}
        \caption{Hour 12}
    \end{subfigure}\hfill
    \begin{subfigure}{0.21\linewidth}
        \includegraphics[width=\linewidth]{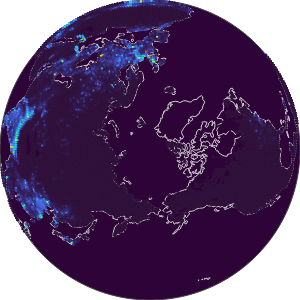}
        \caption{Hour 18}
    \end{subfigure}
    \hfill
    \vspace{-0.3cm}
    \caption{Spatio-temporal model of global Internet demand (viewed from directly above the North Pole with the Sun at the top of the page).}
    \label{fig:spatiotemporal}
\end{figure*}

To illustrate, Figure~\ref{fig:spatiotemporal} shows the time-adjusted bandwidth demand at four distinct points in time through out the day for the entire Northern Hemisphere.
At each point in time we rotate the Earth such that the sun is in the direction of the top of the page.
Though it is somewhat obfuscated by the large absolute differences in population density at different longitudes, clear ``quieter'' and ``louder`` regions are apparent.
(For example, the right-hand side of each figure corresponds to the early hours of the morning and remains dark whereas the top corresponds to midday and remains light.)

{\bf Key takeaway.} A spatiotemporal view of bandwidth demand reveals that Internet usage is neither globally uniform nor temporally flat. 
Rather, demand follows predictable patterns tied to both geography and local time, suggesting that satellite constellations should be engineered not around Earth's surface, but around the sun-relative structure of human activity.

\subsection{Near-Earth radiation}

A secondary driver behind increased numbers of satellites is the need to maintain spare satellites to be swapped in when active satellites fail.

Though satellite failures may originate from a variety of causes, we posit a persistent cause of failure to be exposure to radiation (\ie energetic electrons, protons, and other ions) trapped in the Earth's magnetic field~\cite{guild2022best}.
Although satellites can be built with shielding to protect sensitive electronic components from this radiation, doing so increases both the cost of each satellite and the total mass of material.
Hence reducing aggregate radiation exposure of a satellite constellation reduces the number of spare satellites required for replacing failures as well as the costs of each individual satellite.

\begin{figure}[!htb]
    \centering
    \includegraphics[width=0.8\linewidth]{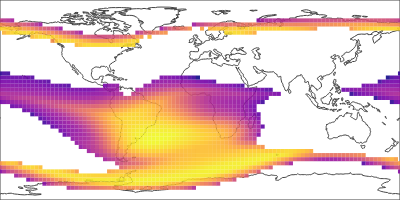}
    \caption{Maximum electron radiation at 560~km altitude over a sample of 128 days from solar cycle 24.}
    \vspace{-0.3cm}
    \label{fig:radiationexample}
\end{figure}

To understand the structure of near-Earth radiation fields, we leverage IRENE---a state-of-the-art dataset and model designed specifically for pre-mission estimate of radiation exposure in Earth orbit~\cite{irene}.
To illustrate, Figure~\ref{fig:radiationexample} shows IRENE's estimate of radiation flux for satellites at 560~km altitude.
Because radiation intensity and structure are strongly dependent on solar activity, we aggregate over a sample of days randomly selected from solar cycle 24.

Several key spatial structures are apparent in Figure~\ref{fig:radiationexample}.
First, a large region of high radiation intensity sits over South America and the South Atlantic---commonly referred to as the South Atlantic Anomaly~\cite{girgis2022radiation}.
Second, distinct bands of high radiation intensity cross over moderate-to-high Southern and Northern latitudes.
These structures result from the intersection of the ``inner'' and ``outer'' Van Allen ``belts'' with the orbital altitude respectively.

\begin{figure}[!htb]
    \centering
    \begin{subfigure}{0.49\linewidth}
        \includegraphics[width=0.9\linewidth]{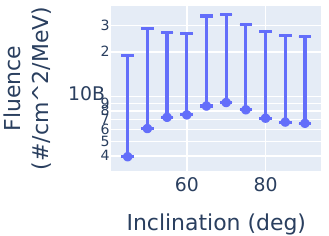}
        \caption{Electrons}
    \end{subfigure}~
    \begin{subfigure}{0.49\linewidth}
        \includegraphics[width=0.9\linewidth]{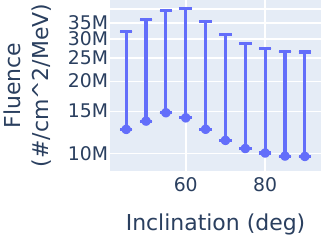}
        \caption{Protons}
    \end{subfigure}
    \caption{Estimated daily radiation exposure for 560~km orbits as a function of inclination.}
    \vspace{-0.3cm}
    \label{fig:radiationVsInclo}
\end{figure}

The key implication of the structure---particularly of the outer Van Allen belt---for LSN design is that the commonly-used moderate inclination orbits (\eg 60$\degree$ to 70$\degree$) actually represent a worst-case scenario from a radiation perspective.
Intuitively, this is because orbital trajectories at this inclination ``turn around'' at their Southern- and Northern-most points directly in these radiation bands, increasing their overall radiation exposure.
In contrast, lower inclination orbits turn around before reaching the radiation band and higher inclination orbits quickly pass through the band to turn around at higher latitudes.
To illustrate this effect, Figure~\ref{fig:radiationVsInclo} shows the average radiation accumulation over a full day for orbits at the same altitude as a function of inclination.

{\bf Key takeaway.} Radiation exposure is not uniform across LEO, and inclination plays a decisive role in determining cumulative satellite damage. 
By selecting orbits that avoid high-radiation regions we can enhance satellite durability and reduce redundancy requirements, thereby lowering both operational risk and environmental cost.

\section{Towards High-efficiency LSN Designs}

The insights into the spatiotemporal nature of bandwidth demand and the structured distribution of radiation exposure warrant a rethinking of LSN design. 
Instead of aiming for geometric uniformity, we believe future constellation designs should exploit the regularity and predictability inherent in both user behavior and orbital physics. 
In this section, we introduce one such design that advocates the use of sun-synchronous orbits, which naturally align with daily demand cycles and reduce radiation exposure, and demonstrate its advantages in both efficiency and suvivability.

\subsection{A sun-synchronous (SS) perspective on demand}

Building on the spatiotemporal model of bandwidth demand discussed in \S~\ref{ssec:structureOfBandwidth}, we propose a ``sun-synchronous'' (SS) perspective on bandwidth demand and LSN constellation design.
The key defining feature of SS orbits is that they pass over any particular latitude at a single, fixed local time of day (\eg always ascending over the equator at noon).
This is achieved by tuning inclination and altitude so that the natural precession of the orbital plane corresponds exactly to the motion of the Earth around the Sun. %
SS orbits require inclinations greater than 90$\degree$ making them ``retrograde'' orbits whose ground-tracks move from East to West (rather than from West to East as is the case for most present-day LSNs).
Though such orbits have higher launch costs (because extra fuel is required to counter the momentum inherited from the spin of the Earth), we anticipate their long-term savings for constellation design will out-weigh these one-time launch costs.

\begin{figure}[!htb]
    \vspace{-0.2cm}
    \centering
    \includegraphics[width=0.9\linewidth]{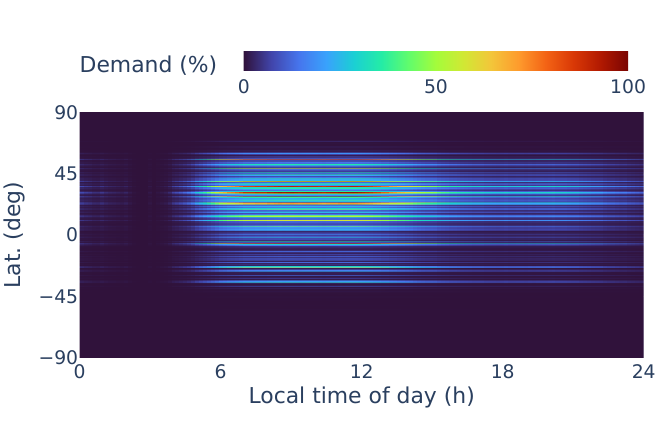}
    \vspace{-0.3cm}
    \caption{Visualization of spatiotemporal structure of demand as a function of the local time-of-day.}
    \label{fig:ssDemand}
    \vspace{-0.2cm}
\end{figure}

To understand the utility of SS orbits for LSN design, consider a non-rotating latitude vs. ``local time of day'' grid fixed w.r.t. the direction of the sun (instead of the usual latitude vs. longitude grid fixed to the Earth's rotating surface).
Each particular latitude, time-of-day point on this grid sees all longitudes as the Earth rotates and hence must be able to provide up to the maximum bandwidth demanded at that latitude scaled for the point's particular (fixed) time of day.
To illustrate, Figure~\ref{fig:ssDemand} shows such a latitude, time-of-day grid estimated from the data discussed in \S~\ref{ssec:structureOfBandwidth}.
The clustering of demand along latitude corresponds to the clustering of population density shown in Figure~\ref{fig:gwpdLat} and the structure of demand along time-of-day corresponds to the temporal structure shown in Figure~\ref{fig:cesnetExample}.

Intuitively, an LSN whose orbital trajectories are fixed to this latitude, time-of-day grid while satisfying the demand at each grid point is also able to satisfy the demand of the (rotating) latitude, longitude grid on the Earth's surface.

\subsection{Covering the demand map}

Our key method is to generate planes of SS orbits which correspond to fixed paths on the latitude vs. time-of-day bandwidth demand grid.
The SS constellation design problem is then a matter of selecting a set of such SS-planes such that the demand at each grid point is satisfied and the number of planes (and hence number of satellites) is minimized.

To efficiently compute approximate solutions to this optimization problem, we develop the following greedy algorithm.
\begin{enumerate}
    \item Select a latitude, time-of-day point that has the maximum bandwidth demand.
    \item Add an SS-plane that intersects that point and subtract the capacity of one satellite from all latitude, time-of-day points covered by the plane's path (clamping to zero).
    \item If all demand is satisfied, return the accumulated set of planes, otherwise go back to (1).
\end{enumerate}
Because each added orbital plane covers a relatively large range of latitude, time-of-day points (in addition to the maximum-demand point), this algorithm quickly generates sets of planes that cover all spatiotemporal demand even though it may not find the minimal number of planes. 

\subsection{Initial performance evaluation}

We evaluate our approach by considering the spatiotemporal bandwidth demand shown in Figure~\ref{fig:ssDemand}.
To avoid dependence on the bandwidth capacity of a particular satellite design or radio technology, we measure bandwidth demand in multiples of the capacity of a single satellite (\ie ``bandwidth multiplier'').
We consider a single orbital altitude ($\sim$560~km) and compare against Walker-delta (WD) constellations constructed by multiple shells (\eg slightly above and below this altitude) at different inclinations determined by maximum population density at each latitude.

\begin{figure}[!htb]
    \centering
    \includegraphics[width=0.9\linewidth]{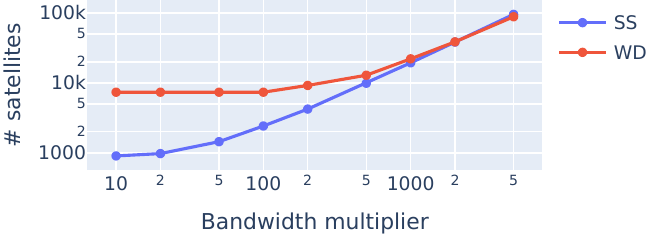}
    \caption{Number of satellites required to satisfy the spatiotemporal structure shown in Figure~\ref{fig:ssDemand} as a function of the total bandwidth demand (measured in multiples of a single satellite's bandwidth capacity).}
    \label{fig:ssNumSats}
\end{figure}

Figure~\ref{fig:ssNumSats} shows the number of satellites required to meet bandwidth demand for both constellation design methods.
At lower bandwidth demands, our approach is able to leverage the spatiotemporal demand non-uniformity whereas the Walker-delta constellation requires more satellites to maintain uniform coverage.
As demand increases, our approach's SS-planes begin saturating the latitude, time-of-day grid leading to a reduction in the difference between methods.

\begin{figure}[!htb]
    \centering
    \begin{subfigure}{0.49\linewidth}
        \includegraphics[width=0.9\linewidth]{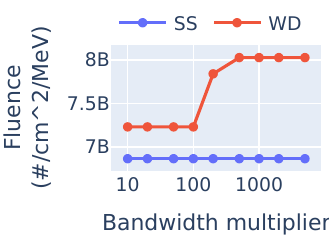}
        \caption{Electrons}
    \end{subfigure}~
    \begin{subfigure}{0.49\linewidth}
        \includegraphics[width=0.9\linewidth]{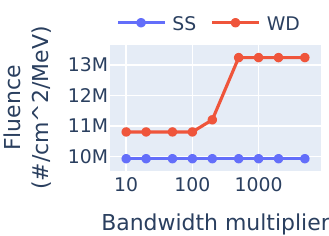}
        \caption{Protons}
    \end{subfigure}
    \caption{Estimated median radiation for a satellite in constellations evaluated in Figure~\ref{fig:ssNumSats}}
    \label{fig:ssRadiation}
\end{figure}

Figure~\ref{fig:ssRadiation} compares the same constellations in terms of their median per-satellite radiation exposure (accumulated over one day using IRENE).
Because all SS-planes have the same inclination, the median radiation exposure does not change as more planes are added.
However, for the Walker-delta constellations, the use of lower inclinations to target higher population densities (see Figure~\ref{fig:gwpdLat}) leads to higher median radiation exposure across the constellation.

{\bf Key Takeaway.} Together, these results demonstrate that sun-synchronous constellations not only require significantly fewer satellites to meet realistic spatiotemporal demand, but also reduce exposure to harmful radiation. Furthermore, these benefits represent a compelling path forward for designing LEO networks that are not only more efficient and survivable, but also more environmentally and economically sustainable.

\section{Implications for Networking}

Traditional networking strategies for LSNs have assumed uniform, always-available infrastructure. 
However, SS-plane constellations bring spatiotemporal awareness to the forefront. 
LSN research must now account for predictable, dynamic coverage patterns that follow the Earth's relation to the sun rather than its rotational geography. This fundamental shift opens up a robust LSN research agenda: %

(1) {\em How can we design time-aware satellite network topologies and routing protocols that proactively exploit predictable spatiotemporal variations in sun-synchronous constellations?} 
First, network topology design must adapt to a constellation layout that is inherently non-uniform in time and space. 
Routing protocols must be capable of handling predictable gaps and surges in connectivity, possibly by precomputing time-aware paths and schedules. 
Moreover, bandwidth allocation and scheduling algorithms should exploit the regularity of human activity to prioritize peak-hour service and shift non-urgent traffic to off-peak periods.

(2) {\em What fault-tolerance and traffic management strategies are best suited to low-failure, low-complexity LEO constellations?}
The reduced failure rate due to lower radiation exposure also alters the survivability paradigm. With fewer random outages, networks can adopt lighter-weight failure recovery mechanisms. 
This opens the door for more efficient use of bandwidth and lower latency in fault detection and re-routing. 
Additionally, the smaller number of satellites simplifies inter-satellite link management, reducing the complexity of coordination and congestion avoidance.

(3) {\em What simulation methodologies and traffic models are required to accurately evaluate LEO network performance in sun-synchronous constellations?} These changes also highlight the need for new simulation tools and modeling frameworks. 
Future research must incorporate sun-relative spatiotemporal traffic models into the design, simulation, and verification of satellite network protocols. 

(4) {\em How should the control plane architecture of satellite networks evolve to incorporate solar-relative demand modeling, predictive resource allocation, and integration with terrestrial networks for seamless end-to-end service delivery?} At a broader architectural level, SS constellations represent a shift toward spatiotemporally optimized infrastructure.
Future network control planes may need to incorporate solar-relative models of demand and availability as first-class primitives. This shift suggests that satellite network controllers, whether centralized or distributed, should evolve to operate on sun-relative coordinates, periodically updated with bandwidth forecasts, satellite health, and user load predictions. 
Integration with terrestrial Internet systems will also need to be revisited, particularly in terms of dynamic handoff between satellites and ground stations, latency-sensitive peering, and caching.

\bibliographystyle{ACM-Reference-Format} 
\bibliography{references}

\end{document}